# A Quantitative Approach to Evaluating Open-Source EHR Systems for Indian Healthcare

Biswanath Dutta[1] and Debanjali Bain[2]
DRTC, Indian Statistical Institute
Bangalore 560059, India
bisu@drtc.isibang.ac.in
debanjali@drtc.isibang.ac.in

**Abstract**. The increasing use of Electronic Health Records (EHR) has emphasized the need for standardization and interoperability in healthcare data management. The Ministry of Health and Family Welfare, Government of India, has introduced the Electronic Health Record Minimum Data Set (EHRMDS) to facilitate uniformity in clinical documentation. However, the compatibility of Open-Source Electronic Health Record Systems (OS-EHRS) with EHRMDS remains largely unexplored. This study conducts a systematic assessment of the alignment between EHRMDS and commonly utilized OS-EHRS to determine the most appropriate system for healthcare environments in India. A quantitative closeness analysis was performed by comparing the metadata elements of EHRMDS with those of 10 selected OS-EHRS. Using crosswalk methodologies based on syntactic and semantic similarity, the study measured the extent of metadata alignment. Results indicate that OpenEMR exhibits the highest compatibility with EHRMDS, covering 73.81% of its metadata elements, while OpenClinic shows the least alignment at 33.33%. Additionally, the analysis identified 47 metadata elements present in OS-EHRS but absent in EHRMDS, suggesting the need for an extended metadata schema. By bridging gaps in clinical metadata, this study contributes to enhancing the interoperability of EHR systems in India. The findings provide valuable insights for healthcare policymakers and organizations seeking to adopt OS-EHRS aligned with national standards.

## 1. Introduction

An electronic health record (EHR) is a digital compilation of data pertaining to a patient's health, well-being, and medical care [1]. It offers a secure and efficient method for storing, sharing, and accessing information, utilizing standardized or widely recognized logical information models. The primary aim of the EHR is to support continuous, efficient, high-quality, and secure integrated healthcare delivery.

The standardized representation of EHR data makes it possible to exchange and communicate patient data efficiently across various healthcare settings. Metadata is of utmost importance in ensuring the efficient and effective management of EHR data. It serves as a well-established approach for the management, maintenance, preservation, and exchange of patients' healthcare data. It enables the identification and retrieval of pertinent information and ensures accurate interpretation and use by healthcare providers. For instance, metadata can be used to identify the date of a patient's last visit, the treatment administered, etc., which allows easy retrieval and interpretation of the relevant information. Metadata enables capturing a patient's record at a detailed level, including individual data elements [2], for example, body temperature, heart rate, respiratory rate, blood pressure, height, weight, etc. The specific data elements within an EHR are crucial for evaluating a patient's overall health status and can offer significant insights into their medical condition. This detailed level of information enables the selective sharing of particular sections of the health record while safeguarding sensitive data. For instance, a patient's medication history can be shared with a new healthcare provider, while more private information, such as mental health records, can remain confidential.

The global adoption of EHR systems is accelerating rapidly. In February 2009, the United States introduced the Health Information Technology for Economic and Clinical Health (HITECH) Act, marking the establishment of the first EHR standard [3]. Similarly, in January 2011, France implemented its initial EHR guideline, known as “Dossier Medical Personnel (DMP)” [4]. Given India's status as the second-most populous country globally, there is a growing demand for high-quality healthcare. In September 2013, the Ministry of Health and Family Welfare (MoHFW), Government of India (GoI) took a significant step towards establishing a consistent and uniform EHR management system in India. To attain this objective, the MoHFW published the Electronic Health Record Standards of the GoI [5]. This publication aimed to provide clear guidelines and standards for hospitals and healthcare providers across the country to follow, ensuring the implementation of a standardized approach to EHR maintenance. These standards include various recommendations, including Electronic Health Record Minimum Data Set (EHRMDS), which aims to encourage the adoption of EHR for capturing, storing, visualizing, presenting, transmitting, and ensuring interoperability in clinical records (for details, see [6]).

Electronic Health Record Systems (EHRS) are software applications (aka EHR tool) that facilitate the creation, storage, and exchange of EHRs among healthcare providers, enabling efficient management of patient health information [7]. Open-Source Electronic Health Record Systems (OS-EHRS) have seen increasing global adoption, with various regions implementing different systems tailored to their specific needs. For instance, OpenEHR is widely used in several Northern European countries, while, GNU Health has gained popularity in diverse countries, including China, the USA, Argentina, Germany, and Spain, demonstrating its versatility and broad applicability across different healthcare contexts. In regions such as Africa, India, and Southeast Asia, OpenMRS has emerged as a highly utilized OS-EHRS. OpenEMR has gained significant utilization in countries, like USA, UK, South Korea, and Brazil [8, 9]. Selecting the most suitable OS-EHRS for a specific healthcare system requires careful consideration of factors such as local requirements, infrastructure, and available resources [10].

The aim of this work is to evaluate the alignment between EHRMDS and OS-EHRS through a quantitative approach, providing a structured and data-driven framework for assessing metadata compatibility. By building upon previous research [6] that qualitatively analyzed the closeness between EHRMDS and OS-EHRS, this study introduces closeness percentage, numerical ranking, and statistical evaluation to objectively measure compatibility. A key objective is to identify gaps in EHRMDS by analyzing additional metadata elements present in OS-EHRS but absent in EHRMDS, leading to the development of an extended metadata schema (EHRMDS-ext). This enriched framework can enhance clinical documentation, interoperability, and data standardization. Furthermore, the study integrates SNOMED-CT and UMLS terminologies to facilitate semantic interoperability and efficient data exchange across healthcare systems. Another key goal is to establish a structured methodology for selecting OS-EHRS, using PRISMA-based literature selection and predefined evaluation criteria such as metadata coverage and system availability. The study also aims to provide actionable insights for policymakers and healthcare organizations by highlighting metadata gaps, offering recommendations for improving standardization and interoperability in Indian healthcare.

In particular, this work aims to answer the following Research Questions (RQs):

- *RQ1: To what extent do Open-Source Electronic Health Record Systems (OS-EHRS) align with the Electronic Health Record Minimum Data Set (EHRMDS) in terms of metadata coverage? To what extent do OS-EHRS align with the EHRMDS in terms of metadata coverage, and what is the average closeness percentage across these systems?*
- *RQ2: Which OS-EHRS exhibits the highest metadata compatibility with EHRMDS, and what percentage of alignment does it achieve?*
- *RQ3: What are the key metadata elements present in OS-EHRS that are missing in EHRMDS, and how do these gaps impact interoperability and data completeness?*
- *RQ4: How does the quantitative closeness analysis help in identifying the best-suited OS-EHRS for the Indian healthcare system?*
- *RQ5: What are the implications of metadata gaps in EHRMDS for standardization and interoperability in electronic health record systems?*

The study provides a comprehensive list of Open-Source EHR tools and analyzes selected OS-EHRS in greater detail. It provides a comprehensive comparison of EHRMDS and OS-EHRS metadata, encompassing their crosswalk and the unified metadata framework for managing electronic health records. Overall, the study aims to assist organizations in making informed decisions about EHR systems and improving the quality of healthcare data. The primary contributions of the present work are:

i. *Systematic Evaluation of Metadata Compatibility: Uses EHRMDS as a benchmark to assess OS-EHRS suitability for the Indian healthcare context. This systematic approach can aid organizations in selecting systems that improve interoperability and patient care.*
ii. *Analytical Framework for Metadata Closeness: Introduces a scalable methodology incorporating metrics such as closeness percentage, gap analysis, and standard deviation. This framework provides a structured and objective way to evaluate metadata alignment, supporting future research and decision-making.*
iii. *Metadata Crosswalk Development: Creates a detailed crosswalk mapping metadata elements between OS-EHRS and EHRMDS. This approach can help organizations identify overlaps and gaps, improving data completeness and interoperability.*
iv. *Quantitative Alignment Assessment: Enhances previous qualitative studies* [6] *by introducing a numerical closeness analysis and percentage-based ranking to evaluate OS-EHRS alignment with EHRMDS. This data-driven approach ensures objective and structured comparisons for selecting the most suitable system.*
v. *Extended Metadata Schema: Identifies missing metadata elements in EHRMDS and develops an extended schema. This enriched framework ensures more comprehensive and standardized clinical documentation, enhancing interoperability.*
vi. *Integration of SNOMED-CT and UMLS: Incorporates international medical terminologies to standardize metadata, enabling better data exchange and interoperability across healthcare systems. This alignment ensures compatibility with globally accepted medical vocabularies.*
vii. *Structured OS-EHRS Selection Methodology: Introduces a systematic approach using a PRISMA-based literature review and evaluation criteria such as literature frequency and online demo availability. This methodology ensures a transparent and well-informed selection process tailored to Indian healthcare needs.*
viii. *Policy and Organizational Insights: Provides a structured framework for OS-EHRS selection based on metadata coverage and highlights gaps in EHRMDS. The findings offer actionable recommendations for policymakers and healthcare organizations to improve standardization and interoperability, enhancing the effectiveness of electronic health records in India*.

The rest of this paper is structured as follows: Section 2 provides an overview of the background, discussing the Electronic Health Record Minimum Data Set (EHRMDS) and Open-Source Electronic Health Record Systems (OS-EHRS). Section 3 reviews related work on metadata crosswalks and interoperability in healthcare systems. Section 4 outlines the methodology, detailing the identification and selection of vocabularies, the closeness analysis approach, and the development of an extended metadata schema. Section 5 presents the results of the quantitative closeness analysis, highlighting the alignment between EHRMDS and OS-EHRS. Section 6 discusses key findings, including metadata gaps and implications for interoperability. Finally, Section 7 concludes the paper, summarizing the contributions and suggesting directions for future research.

## 2. Background

### *2.1. EHRMDS*

In September 2013, the Indian government's Ministry of Health and Family Welfare published "Recommendations on standards of electronic medical records in India" [11]. It aims to establish a standard for EHR that would enable efficient capture, storage and retrieval, visualization, and exchange of healthcare information. Electronic Health Record Minimum Data Set (EHRMDS) was released as part of the stated recommendations. The EHRMDS is a set of minimum data elements that are deemed essential in EHR systems for the purpose of efficient capture, storage and exchange of healthcare-related information during clinical encounters.

EHRMDS primarily evolved from the CCR, or Continuity of Care Record [12], a standardized specification developed collaboratively by ASTM International, the Massachusetts Medical Society (MMS), the Healthcare Information and Management Systems Society (HIMSS), the American Academy of Family Physicians (AAFP), the American Academy of Pediatrics (AAP), and other health informatics vendors. The CCR standard is designed to offer a concise summary of a patient's health information, facilitating seamless sharing and comprehension among healthcare providers, patients, and other authorized parties.

As per the above-mentioned recommendations, an EHRS must include all the minimum essential elements indicated in the EHRMDS. Additionally, healthcare providers have the flexibility to incorporate additional elements based on their specific clinical requirements. The EHRMDS pertains a set of 91 elements that encompass different aspects of health data, including patient demographics, insurance information, diagnostic report, medications, vital signs, allergies, and healthcare reports. These elements are classified into ten categories based on their types as shown in Table 1. The table provides a summary of the number of elements within each category along with some example elements. For instance, the “Demographics” category of EHRMDS recommends the capture of a patient's name, age, gender, address, phone number, and email address. The “clinical examination” category captures the details of the patient's physical examination during a clinical encounter. This also includes information on the patient's vital signs, such as heart rate, blood pressure, respiratory rate, body temperature, and oxygen saturation. Similarly, “Encounter” captures mandatory data elements, such as date and time of the encounter and the chief complaint. “Medication” captures the data, such as the medication name, dose, frequency, and start and end dates.

**Table 1.** Categorized elements of EHRMDS

| Sl. No. | Category | Description | # of Elements | Example Elements |
|---|---|---|---|---|
| 1 | Identifiers | Include the identity of the entity. | 3 | UHID, Alternate UHID, Insurance ID |
| 2 | Demographics | Include identifying information. | 42 | Patient name, Age, Address |
| 3 | Status | Establishes the state of particulars. | 3 | Organ Donor Status, Insurance Status, Allergy Status |
| 4 | Episode | A distinctive healthcare event. | 2 | Episode type, Episode Number |
| 5 | Encounter | A casual healthcare contact between patient and healthcare provider. | 4 | Encounter Type, Encounter Date & Time, Reason for Visit |
| 6 | History | The aggregate of occurred or ongoing medical events. | 8 | Present History, Personal History, Immunization History, Allergy History |
| 7 | Clinical Examination | Establishes the nature, implications, and result of the clinical findings. | 13 | Clinical Exam Vitals Systolic BP, Clinical Exam Pulse Rate, Clinical Exam Temperature (°C), Clinical Exam Height (cms) |
| 8 | Diagnosis | A decision on the clinical condition identifying the nature or cause. | 4 | Diagnosis Type, Diagnosis (Description) |
| 9 | Treatment Plan | A detailed plan on the patient’s disease, goal, treatment options, and duration. | 6 | Treatment Plan Investigations, Treatment Plan Medication, Treatment Plan Procedure, Treatment Plan Referral |
| 10 | Medication | For alleviating or treating the illness with medicine. | 6 | Medication Name, Strength, Dose, Route, Frequency |

### 2.2. *EHR Systems*

Electronic Health Record Systems (EHRS), also known as EHR tools, are software applications or digital platforms designed to store, manage, and exchange patient healthcare-related information electronically in a secure and standardized manner. They aim to create a centralized repository of patient related data, allowing healthcare professionals to access the information for clinical decisions making [7]. EHRS can offer comprehensive features such as e-prescribing, billing, appointment scheduling, electronic health record-keeping, seamless management of administrative tasks, tracking patient encounters, managing referrals, and facilitating effective care coordination among healthcare professionals [10]. These functionalities improve workflow efficiency, reduce manual tasks, enhance communication.

The EHR systems have been purposefully designed to capture and integrate various types of health information on patient’s health status. EHRS can serve various functions to support healthcare workflows and patient care. EHRS can offer a portal that empowers individuals to access their health records, schedule visits or appointments, and communicate with healthcare providers [13]. EHRS can facilitate the capture and storage of structured, semi-structured and unstructured patient data from multiple resources (such as a paper-based handwritten doctor’s note, or prescription). These systems are essential in enabling smooth communication and collaboration among various healthcare stakeholders, leading to more coordinated and effective care. With Electronic Health Record Systems (EHRS), healthcare providers can effortlessly share patient health information, eliminating the need for emailing or mailing physical records [14]. This enhances efficiency, as providers can promptly access the necessary data to make well-informed decisions about patient care. Some popular examples of EHRS include OpenEHR, OpenEMR, GNUMed, OpenMRS, OSCAR, etc.

## 3. Related Work

This section presents studies focused on determining the resemblance between EHR metadata elements from various EHR systems and EHR metadata standards. Additionally, this section explores different methods of overlapping and cross-walking between EHR standards-specific metadata elements.

Chen et al. (2009) focused on comparing metadata in the EHR standard OpenEHR and the Swedish EHR system Cambio COSMIC [15]. A conceptual mapping was conducted between the COSMIC, RM and OpenEHR AM. The study indicated notable resemblances between OpenEHR Archetype model and the COSMIC model.

Ferranti et al. (2006) conducted a critical analysis of two electronic health record standards: CCR of the ASTM International and CDA of HL7 [16]. The CCR is a concise collection of healthcare data that encompasses information from healthcare professionals, insurance providers, and medical information, such as pharmaceuticals, diagnostics, vital signs, symptoms of allergy, outcomes, recent clinical procedures, and more, whereas the CDA is commonly used for radiology findings, clinical summaries, reports on progress, and discharge summaries. Ferranti et al. suggested a method for integrating CCR and CDA by describing a collection of distinctive metadata items based on the information in both standards.

Müller et al. (2005) focused on the development of HIS to facilitate healthcare-related data transfer [17]. The system was designed based on the widely used CDA. To ensure seamless interoperability between the HIS and CDA, a mapping was carried out, which allowed the researchers to establish a clear correspondence between the elements of CDA and the corresponding terms in the HIS.

The HL7 International Electronic Health Record Technical Committee (Dolin et al., 2006) conducted a thorough crosswalk analysis [18]. The researchers focused on evaluating the CDA Header, Lifecycle Modelling, and RM-ES Profile criteria. The aim was to ascertain metadata terms that exhibit close association across various frameworks. To improve the harmonization and interoperability of EHRS, the committee proposes a conceptual overlap between the CDA R2 and the Interoperability Model.

Cucchiara (2014) focused on generating a crosswalk and alignment between two significant frameworks in healthcare: PCMH model (introduced by the American Academy of Pediatrics) and MU criteria [19]. The aim was to identify areas of overlap and alignment between these two frameworks. Cucchiara successfully generated a comprehensive crosswalk and alignment that highlighted the commonalities and shared elements between these frameworks. The findings of Cucchiara's study concluded that there were indeed numerous areas of overlap between PCMH and MU.

Building upon this work, Coffin et al. (2014) delved deeper into the intersection and crosswalk between specific MU criteria and PCMH requirements [20]. Their study aimed to provide a detailed understanding of how the specific criteria outlined in the Meaningful Use framework can effectively meet the requirements of the Patient-Centered Medical Home model. By identifying the points of convergence and mapping the relevant MU criteria to the corresponding PCMH requirements, Coffin et al. shed light on the practical application and alignment between these two frameworks.

From the above discussion, we can observe that none of the existing studies investigated the closeness of EHRMDS India to the OS-EHRS. Besides, despite the prevalence of metadata crosswalks in healthcare and other domains, there has been no work to quantify the closeness between metadata vocabularies. The current study fills these gaps and provides valuable insights into the similarities between the EHRMDS India and the OS-EHRS. Table 2 provides a summary of the related works described above.

**Table 2.** Summary of the related works on EHR Metadata Crosswalks and Interoperability

| Authors [Ref.] | Purpose | Vocabulary used for the study | Final Output | Quantifiable Closeness Analysis? | Production of a uniform extended list? |
|---|---|---|---|---|---|
| Chen et al. [15] | Study similarity between EHR system and standard | Cambio COSMIC and OpenEHR | Semantic mapping between RM and AM of OpenEHR and COSMIC. | Yes | No |
| Ferranti et al. [16] | Critically evaluate EHR standards | CDA and CCR | Integrating CDA and CCR by defining a set of common data elements. | No | Yes |
| Muller et al. [17] | Develop HIS for electronic data transfer | CDA and HIS | Mapped CDA elements to corresponding HIS terms. | No | No |
| HL7 International EHR Technical Committee [18] | Crosswalk between key criteria to identify related metadata terms | CDA R2 Header, Lifecycle Model, and RM-ES Profile | Single list of metadata concepts and term definitions with overlap of concepts between Interoperability Model and CDA R2. | No | Yes |
| Cucchiara [19] | Generate a crosswalk and alignment between two models | PCMH and MU | Crosswalk and alignment between PCMH and MU with areas of overlap identified. | No | No |
| Coffin, et al. [20] | Investigate the alignment of EHRMDS India to OS-EHRS | MU criteria and PCMH requirements | Intersection explain how specific MU criteria can meet PCMH requirements | No | No |

## 4. Methodology

The methodology follows a structured approach divided into three main phases (as illustrated with blue coloured filled), each comprising specific steps (as illustrated with green coloured filled) to achieve the study's aim, as depicted in Figure 1.

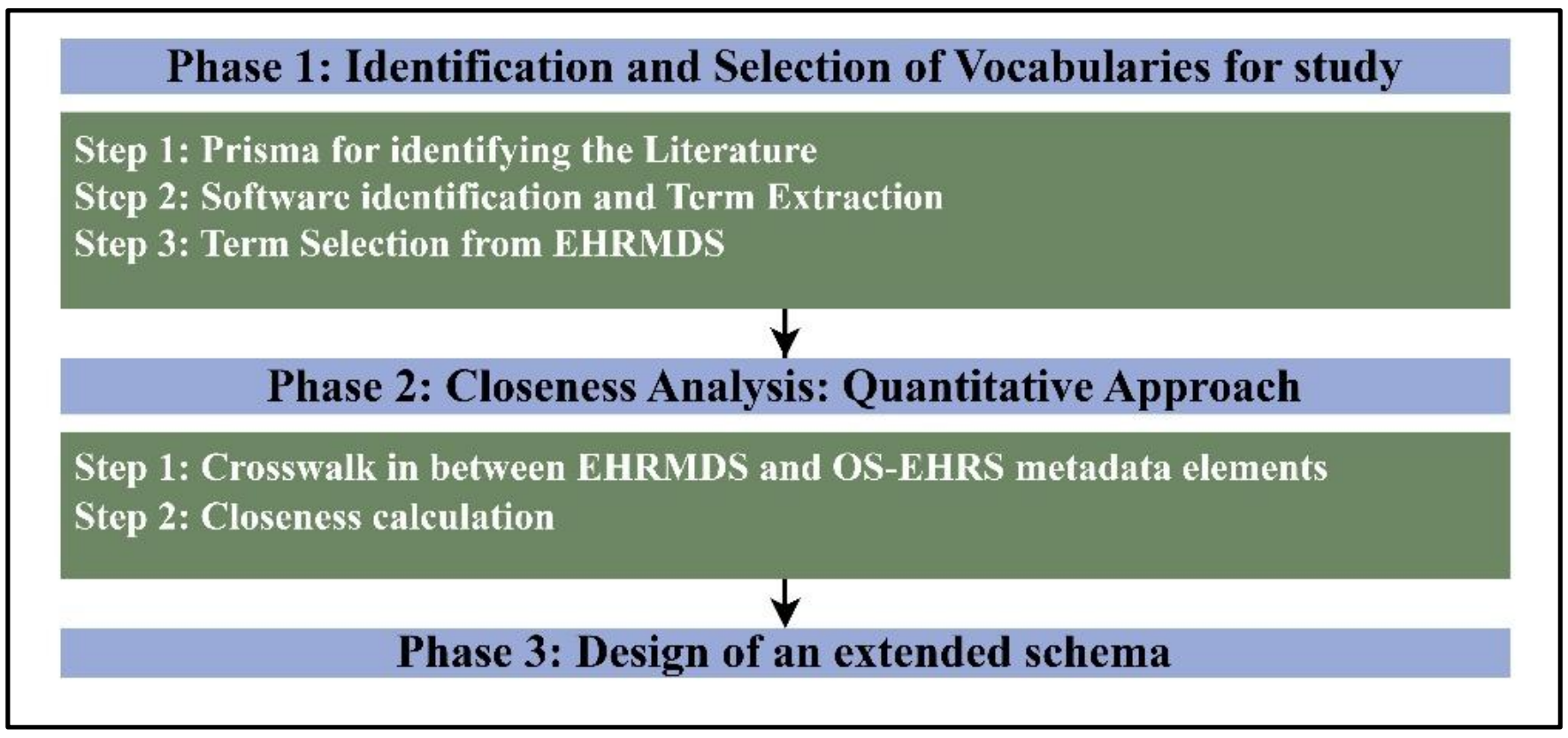

**Figure 1.** Methodology overview.

### 4.1. *Phase 1: Identification and Selection of Vocabularies for study*

This phase focuses on selecting relevant vocabularies that are essential for the research. The following steps are carried out:

*Step 1: PRISMA for Identifying Literature* – A systematic review methodology (PRISMA) [21] was applied to identify relevant literature and datasets, ensuring comprehensive coverage of existing knowledge.

*Step 2: Software Identification and Term Extraction* – Step 2 involves EHRS identification and term extraction.

*Step 3: Term Selection from EHRMDS* – Step 3 involves the selection of terms from our benchmark vocabulary, i.e., EHRMDS.

### *4.2. Phase 2: Closeness Analysis – Quantitative Approach*

This phase involves a quantitative assessment to determine the closeness of metadata elements between different schemas. The following steps were performed:

*Step 1: Crosswalk in between EHRMDS and OS-EHRS metadata elements* – A crosswalk analysis was conducted to map corresponding metadata elements between the EHRMDS and OS-EHRS. This allows the comparison of elements across systems.

*Step 2: Closeness Calculation* – A quantitative measure of similarity was calculated between the metadata elements to assess their alignment and compatibility. In this step, the closeness of each software system to the benchmark EHRMDS is calculated by counting the percentage of EHRMDS elements that match each EHRS. This analysis helps identify the EHRS that is most suitable for India in consideration of healthcare metadata coverage.

### *4.3. Phase 3: Design of an Extended Schema*

The third phase of the study encompasses the development of extended clinical metadata. This schema was developed by integrating elements of different EHRS and expanding the existing EHRMDS. The schema is designed to improve interoperability and standardization within electronic health record systems.

By following this structured methodological approach, the study ensures a comprehensive analysis and formulation of an improved metadata schema.

## 5. Results

### *5.1. Phase 1: Identification and Selection of Vocabularies for study*

The process of identifying and selecting OS-EHRS has been carried out in three distinct steps, as provided below.

*Step 1: PRISMA for identifying the Literature*
The first step involved a comprehensive literature search following the PRISMA flowchart shown in Figure 2. It was conducted to retrieve scholarly publications about OS-EHRS. The search was carried out across several prominent databases, including PubMed[1], ScienceDirect[2], Springer[3], IEEE Xplore[4], and Google Scholar[5].

[1] https://pubmed.ncbi.nlm.nih.gov/
[2] https://www.sciencedirect.com/
[3] https://www.springer.com/
[4] https://ieeexplore.ieee.org/
[5] https://scholar.google.com/

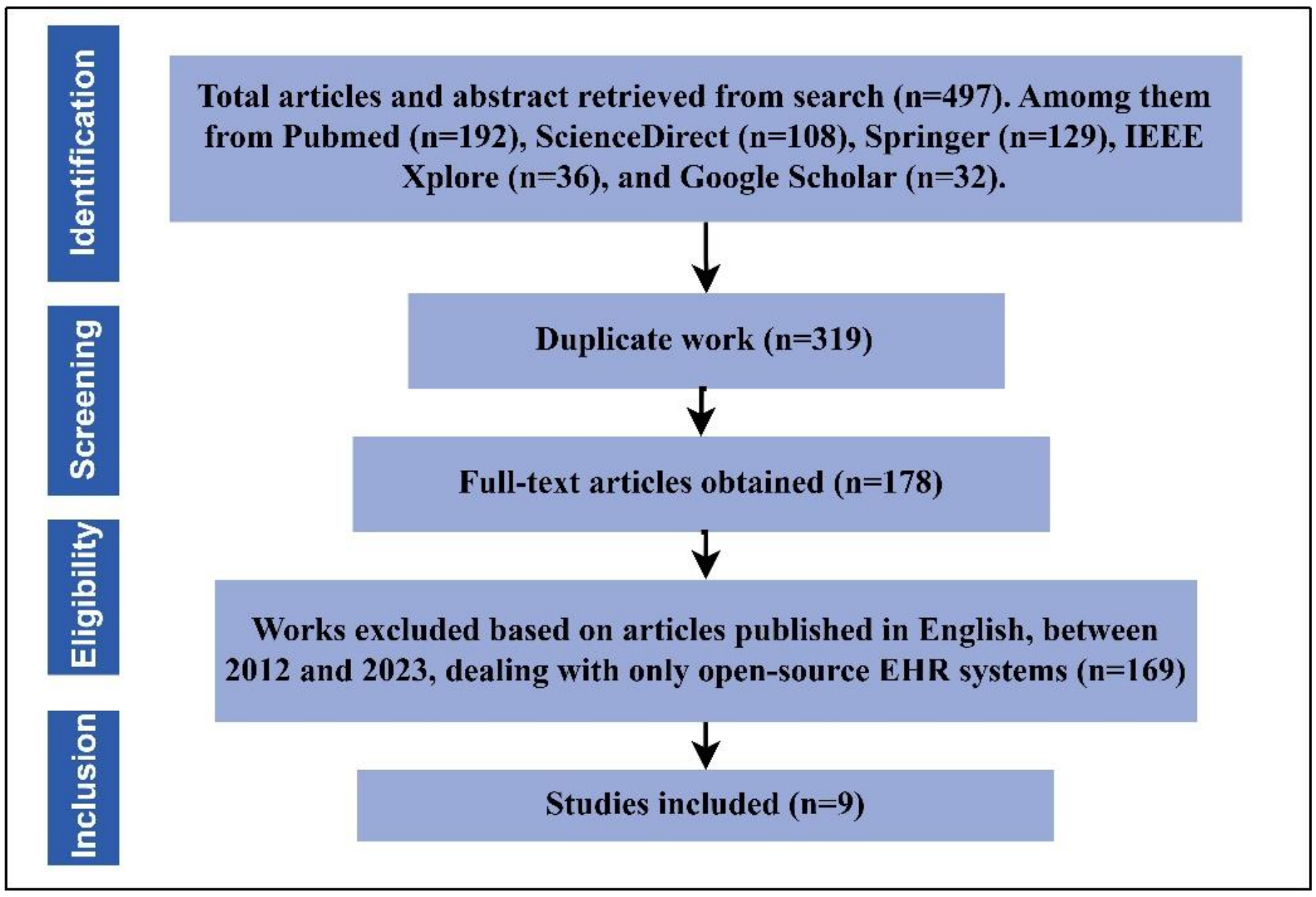

**Figure 2.** Systematic search process used to identify and select relevant literature.

The articles gathered from the databases were screened using targeted keywords, including "electronic health records", "computerized patient records", "digital health records", "open-source EHR systems", "comparison of open-source EHR platforms", "evaluation of open-source EHR solutions", "top electronic health record systems", "free EHR software", and "rankings of open-source EHR systems". Only articles that were written in English, published between 2012 and 2023, and discussed open-source EHR systems were considered for further analysis.

Initially, the literature search yielded a total of 497 publications. After removing duplicates and filtering out irrelevant articles, the number was reduced to 319. Out of these, we could only access the full-text literature for 178 articles. Subsequently, 169 works were excluded based on the selection criteria, leaving us with a set of nine core literature that were used for the identification of EHRS. The details of this core literature are provided in Table 3.

**Table 3.** EHRS referred in selected literature

| Sl. No. | Authors [Ref.] | Description | Tools Referred |
|---|---|---|---|
| 1 | Flores & Vergara [7] | Studied the functionalities of free and open-source EHRs. | CHITS, GNUmed, Open-EMR, OpenMRS, OSCAR, and PatientOS |
| 2 | Multak et al. [22] | Evaluated multiple EMR systems by considering acceptance in healthcare, inpatient & outpatient support, community support, and update frequency. | OpenVista, WorldVistA, Astronaut, ClearHealth, VistaA, WebVista, OpenMRS, Care2X, OpenEMR, OSCAR, Patient OS, GNUHealth, GNUmed, THIRRA, FreeMED |
| 3 | de la Torre et al. [23] | Analyzes open-source EHRs based on specific criteria. | HOSxp, OpenEMR, and OpenVistaA |
| 4 | Kiah et al. [24] | Analyzes available open-source EHRs. | FreeMED, GNUmed, OSCAR, GNU Health, Hospital OS, Solismed, OpenEMR, THIRRA, OpenMRS, WorldVistA, ZEPRS, ClearHealth, MedinTux |
| 5 | Zaidan et al. [25] | Evaluates open-source EHRs based on a set of criteria. | GNUmed, OpenEMR, and OpenMRS ZEPRS |
| 6 | Mo, D. [26] | Discusses the top 26 free and open-source EMR-EHR systems for Windows, Linux, and Mac OSX. | HospitalRun, Open-MRS, Bahmni, FreeMed, OpenEMR, Cottage Med, GNU med, Open-Clinic, OpenEyes, World-VistA, OpenMRS, GNU Health, FreeMed-Forms, ZEPRS, SMART pediatric Growth, OpenHospital, Libre-HealthEHR, THIRRA, FreeHealth.io, Medin-Tux, DolibMed EMR, NoshEMR, DOQ EMR, Chikitsa |
| 7 | Purkayastha et al. [9] | Identifies the most popular OS-EHRS based on Alexa web ranking and Google trends. | OSHERA VistA, GNU Health, Open Medical Record System (Open-MRS), Open Electronic Medical Record (Open-EMR), and OpenEHR |
| 8 | Kumar, R. [27] | Analyzes and lists three best open-source EHR solutions listed on Capterra. | 75Health, OpenEMR, OpenMRS |

| Sl. No. | Authors [Ref.] | Description | Tools Referred |
|---|---|---|---|
| 9 | Hedges, L. [28] | Analyzes and compares best free and open-source EHRs. | TalkEHR, 75Health, OpenEMR, One-TouchEMR, OpenMRS |

*Step 2: Software identification and Term extraction*

From the above mentioned nine literature (see Table 3), we have identified a total of 70 EHRS. Once the duplicates were excluded, the number of OS-EHRS identified was 40.

In this step, we aimed to select the most suitable OS-EHRS for our research. Given the vast number of systems available, it was beyond the scope of our work to study the metadata elements of all 40 OS-EHRS. Therefore, we needed to identify a set of criteria for selecting the EHRS that would be most relevant to our study. We decided to apply two key criteria: Literature-Based Frequency (LBF) and Online Accessibility for Evaluation (OAE).

Using LBF, we were able to assess the user acceptability and popularity of the EHRS (see Table 4). We calculated the LBF of each OS-EHRS in the literature that we had collected in Step 1. In the table if the specific OS-EHRS available in the literature (Sl. No. 1-9), the "√" marks in grey box is given, if not available the "×" in white box is given. We calculated the LBF total score for each tool based on their frequency of occurrence. The systems that were mentioned most frequently in the literature were considered to be the most popular and were therefore selected for further investigation.

Another key criterion for our selection was OAE. We considered the availability of online accessibility an essential criterion for selection. By providing a demo, the vendors allow potential users to explore the system's features, usability, and workflow. Therefore, we evaluated the availability of an online demo for each tool before making the final selection. As a result, we excluded certain tools (such as OpenEHR, OpenVista, Bahmni, etc., as in Table 5) that were popular but did not have an online demo system. In this manner, out of a total of 40 OS-EHRS, we excluded 30 and included 10 in our study (illustrated in Table 5).

**Table 4.** Literature-Based Frequency **(**LBF)

| Sl. No. | F/OSS EHRs [Website] | References Sl.No. from Table No. 3 | | | | | | | | | LBF Total |
|---|---|---|---|---|---|---|---|---|---|---|---|
| | | 1 | 2 | 3 | 4 | 5 | 6 | 7 | 8 | 9 | |
| 1 | OpenEMR [www.open-emr.org] | √ | √ | √ | √ | √ | √ | √ | √ | √ | 9/9 |
| 2 | OpenMRS [www.openmrs.org] | √ | √ | × | √ | √ | √ | √ | √ | √ | 8/9 |
| 3 | GNUmed [www.gnumed.org] | √ | √ | × | √ | √ | √ | × | × | × | 5/9 |
| 4 | GNU Health [ftp.gnu.org] | × | √ | × | √ | × | √ | √ | × | × | 4/9 |
| 5 | FreeHealth [www.freehealth.io] | × | √ | | √ | × | √ | × | × | × | 3/9 |
| 6 | HospitalRun [https://hospitalrun.io] | × | × | √ | √ | × | √ | × | × | × | 3/9 |
| 7 | OSCAR [http://oscar-emr.com] | √ | √ | × | √ | × | × | × | × | × | 3/9 |
| 8 | THIRRA [http://thirra.primacare.org.my/] | × | √ | × | √ | × | √ | × | × | × | 3/9 |
| 9 | WorldVistA [https://worldvista.org] | × | √ | × | √ | × | √ | × | × | × | 3/9 |
| 10 | 75Health [www.75health.com] | × | × | × | × | × | × | × | √ | √ | 2/9 |
| 11 | ClearHealth [http://clear-health.com] | × | √ | × | √ | × | × | × | × | × | 2/9 |
| 12 | MedinTux [www.medintux.org] | × | × | × | √ | × | √ | × | × | × | 2/9 |
| 13 | OpenVista [www.open-vista.com] | × | √ | √ | × | × | × | × | × | × | 2/9 |
| 14 | PatientOS [https://sourceforge.net/projects/patientos/] | √ | √ | × | × | × | × | × | × | × | 2/9 |
| 15 | ZEPRS [https://github.com/chrisekelley/zeprs] | × | × | × | √ | × | √ | × | × | × | 2/9 |
| 16 | Astronaut [http://astronautix.com] | × | √ | × | × | × | × | × | × | × | 1/9 |
| 17 | Bahmni [https://www.bahmni.org] | × | × | × | × | × | √ | × | × | × | 1/9 |
| 18 | Care2X [https://care2x.org/] | × | √ | × | × | × | × | × | × | × | 1/9 |
| 19 | Chikitsa [https://chikitsa.net] | × | × | × | × | × | √ | × | × | × | 1/9 |
| 20 | CHITS [https://www.chits.org/] | √ | × | × | × | × | × | × | × | × | 1/9 |
| 21 | Cottage Med [https://cottagemed.org/p4/Cottage-Med] | × | × | × | × | × | √ | × | × | × | 1/9 |
| 22 | DoliMed EMR [https://www.dolimed.com/] | × | × | × | × | × | √ | × | × | × | 1/9 |
| 23 | ERPNext [https://erpnext.com/] | × | × | × | × | × | √ | × | × | × | 1/9 |

| Sl. No. | F/OSS EHRs [Website] | References Sl.No. from Table No. 3 | | | | | | | | | LBF Total |
|---|---|---|---|---|---|---|---|---|---|---|---|
| | | 1 | 2 | 3 | 4 | 5 | 6 | 7 | 8 | 9 | |
| 24 | FreeMedForm [www.freemedform.com] | × | × | × | × | × | √ | × | × | × | 1/9 |
| 25 | HospitalOS [https://www.hospital-os.com/] | × | × | × | √ | × | × | × | × | × | 1/9 |
| 26 | LibreHealth EHR [https://librehealth.io/projects/lh-her] | × | × | × | × | × | √ | × | × | × | 1/9 |
| 27 | NoshEMR [https://noshemr.org] | × | × | × | × | × | √ | × | × | × | 1/9 |
| 28 | ODOO EMR [https://www.odoo.com] | × | × | × | × | × | √ | × | × | × | 1/9 |
| 29 | OneTouchEMR [www.onetouchemr.com] | × | × | × | × | × | × | × | × | √ | 1/9 |
| 30 | OpenClinic [https://openclinic.sourceforge.net/] | × | × | × | × | × | √ | × | × | × | 1/9 |
| 31 | OpenEHR [https://www.openehr.org] | × | × | × | × | × | × | √ | × | × | 1/9 |
| 32 | OpenEYES [https://theopeneyes.org] | × | × | × | × | × | √ | × | × | × | 1/9 |
| 33 | OpenHospital [https://www.open-hospital.org/] | × | × | × | × | × | √ | × | × | × | 1/9 |
| 34 | OpenMAXIMS [https://github.com/IMS-MAXIMS/openMAXIMS] | × | × | × | × | × | √ | × | × | × | 1/9 |
| 35 | OSHERAVistA [https://worldvista.org/AboutVisA] | × | × | × | × | × | × | √ | × | × | 1/9 |
| 36 | SMART Pediatric Charts [https://github.com/smart-on-fhir/smart-chart-gap] | × | × | × | × | × | √ | × | × | × | 1/9 |
| 37 | Solismed [www.solismed.com] | × | × | × | √ | × | × | × | × | × | 1/9 |
| 38 | TalkEHR [https://www.talkehr.com] | × | × | × | × | × | × | × | × | √ | 1/9 |
| 39 | VxVista [https://www.vxvista.org] | × | √ | × | × | × | × | × | × | × | 1/9 |
| 40 | WebVistA [https://webvista.org] | × | √ | × | × | × | × | × | × | × | 1/9 |

**Table 5.** Analysis of ten OS-EHRS (this table also shows Online Accessibility for Evaluation (OAE) for each OS-EHRS)

| Tool Ref. No. | Software | URI | Type | Programming Language | Database | Platform | Access | Licenses | Standard | Latest Version | Last Release Date | OAE (i.e., demo availability, online access) |
|---|---|---|---|---|---|---|---|---|---|---|---|---|
| T1 | 75Health | www.75health.com | Ambulatory | Python | MySQL | Cross-platform | Web-based | GPL | HL7 CCOW (clinical context object workgroup) standard | 1.0.29 | Jan 13, 2021 | √ |
| T2 | OpenEMR | www.open-emr.org | Ambulatory | PHP | MySQL | Cross-platform | Web-based | GPL | OpenEmrOri ginalSchema | 6.0.0 | May 1, 2021 | √ |
| T3 | OpenMRS | www.openmrs.org | Ambulatory | JAVA | MySQL | Cross-platform | Web-based | GPL | Customizable | 2.9.0 | April 8, 2019 | √ |
| T4 | Solismed | www.solismed.com | Ambulatory | Python | MySQL | Cross-platform | Web-based | GPL | Not Defined | 5.2 | July, 2023 | √ |
| T5 | GNUmed | www.gnumed.org | Ambulatory | Unix Shell, Python, PL/SQL | Python Database API, PostgreSQL | Cross-platform | Server/Client | GPL | ICPM,ICD 9,PCS | Client 1.8.3, Server 22.13 | July 23, 2020 | √ |
| T6 | NoshEMR | www.nosh-emr.org | Ambulatory | Laravel5 PHP | MySQL | Cross-platform | Web-based | GPL | Not Defined | Unknown | Unknown | √ |
| T7 | Freehealth | www.freehealth.io | Ambulatory | PHP | MySQL | Cross-platform | Web-based | GPL | ATC, ICD-10 | 0.11.0 | Dec 30, 2017 | √ |
| T8 | GNUHealth | ftp.gnu.org | Ambulatory/Inpatient | Python | PostgreSQL | Cross-platform | Web-based | GPL | ICPM, ICD10 PCS, ICD9 Vol 3 etc. | 3.8.0 | Feb 14, 2022 | √ |
| T9 | OneTouch EMR | www.onetouchemr.com | Ambulatory | JAVA | MySQL | Cross-platform | Web-based | GPL | Not found | 1.6.19 | Aug 24, 2019 | √ |
| T10 | OpenClinic | openclinic.sourceforge.net | Ambulatory/Inpatient | Python | MySQL | Cross-platform | Web-based | GPL | ICD-10, ICPC-2, Snomed CT, HL7/FHIR API | 0.7 | Mar 11, 2010 | √ |

Table 6 lists the selected OS-EHRS along with their corresponding number of metadata elements available for representing clinical information. For instance, 75Health (T1) offers 48 elements, while OpenEMR (T2) provides 41 elements. For a comprehensive list of the elements available in each tool, see Table 10. The authors manually extracted these elements by visiting the systems and analyzing their metadata.

**Table 6.** Count of metadata elements for each selected OS-EHRS

| OS-EHRS | 75Health (T1) | Open EMR (T2) | OpenMRS (T3) | Solismed (T4) | GNUMed (T5) | NoshEMR (T6) | Freehealth (T7) | GNUHealth (T8) | Onetouchemr (T9) | Openclinic (T10) |
|---|---|---|---|---|---|---|---|---|---|---|
| **# of elements** | 48 | 41 | 28 | 49 | 26 | 38 | 28 | 31 | 38 | 21 |

*Step 3: Term selection from EHRMDS*

The third step entailed choosing the essential EHRMDS metadata elements that are related to clinical data. The EHRMDS consists of a total of 91 metadata elements. In our study, we conducted a thorough analysis and selected a total of 42 elements for evaluation, which are listed in Table 10. We excluded 49 elements, specific to demographic information (such as, the patient's name, patient's age, patient's gender, patient's address, patient's telephone number, etc.).These fields were considered insignificant for our study since they are present in all EHR systems. We included all the mandatory elements from the EHRMDS that are required to be included in any EHR tool. For instance, "Vital signs" and "Immunization" were included as they are critical clinical data elements that play a crucial role in clinical decision-making.

### 5.2. *Phase 2: Closeness Analysis: Quantitative Approach*

In this phase, we examine the correlation among the OS-EHRS and the EHRMDS. This phase comprises two distinct steps, as elaborated below.

*Step 1: Crosswalk in between EHRMDS and OS-EHRS metadata elements*

This preliminary step entails the creation of a crosswalk among the metadata elements of EHRMDS and OS-EHRS. We extract the elements from both EHRMDS and OS-EHRS to initiate this procedure. In Step 3 of Phase 1, the OS-EHRS metadata elements were already extracted. As for EHRMDS, we utilize it as a reference model. Upon obtaining the metadata elements from both EHRMDS and OS-EHRS, a crosswalk is executed. The crosswalk is a process where we compare the metadata elements of OS-EHRS with those of EHRMDS to study their similarity. The comparison is conducted following the syntactic and semantic analysis .The syntactic analysis focuses on the structure and form of terms, disregarding their meaning. It examines the visual or lexicographical similarity between terms. For instance, terms like "BP", "blood pressure" and "B/P" are considered syntactically similar due to their shared structure and abbreviation. On the other hand, semantic analysis aims to identify terms that have the same or similar meanings, regardless of their visual or lexicographical differences. For example, terms like "Medication" and "Drug" are considered semantically equivalent, as they both refer to the concept of medication or drugs in the context of EHR.

During crosswalk conducted in an Excel spreadsheet, we utilized a mathematical intersection approach [29] to establish mappings between different data elements. This mapping process can be represented using the following formula:

$$\bigcap_{\{i=0\}}^{n} Tool_i = Tool_0 \cap Tool_1 \cap Tool_2 \cap \cdots \cap Tool_{10}$$

Here, $Tool_0$ represents data elements of EHRMDS, while $Tool_1$ to $Tool_{10}$ represent data elements of ten different OS-EHRS. Mapping process takes into account both syntactic similarities and semantic equivalences between the elements. For example, let's consider the intersection between $Tool_1$ and $Tool_2$, denoted as ($Tool_1 \cap Tool_2$). This intersection comprises the elements that are present in both $Tool_1$ and $Tool_2$. Intersection between set of terms for $Tool_1$ and $Tool_2$ can be denoted as *($Tool_1$) ∩ ($Tool_2$) = {Term-W, Term-X, Term-Y} ∩ {Term-X, Term-Y, Term-Z} = {Term-X, Term-Y}.*

In this example, 'Term-X" and "Term-Y" are included in the mapping as they exhibit both syntactic similarity and semantic equivalence. The mapping process was performed for all relevant EHR data elements, and detailed results can be found in the corresponding mapping Table 10.

*Step 2: Calculating closeness*

After the completion of crosswalk, next step involves calculating the alignment of EHRMDS to each OS-EHRS. The objective is to identify the most appropriate EHR system for India based on its healthcare-related metadata coverage, according to requirements by EHRMDS. The closeness is determined by calculating the proportion of matching EHRMDS elements in each EHR system. This helps to identify which EHR system is most aligned with the EHRMDS. The formula for closeness analysis is –

$$C(T_i) = \left(\frac{NEHRMDS(T_i)}{N}\right) \times 100\%$$

Where,

$C(T_i)$: represents the closeness of OS-EHRS Ti to EHRMDS.

$NEHRMDS(T_i)$: represents the number of EHRMDS elements that match with OS-EHRS Ti.

$N$: represents the total number of EHRMDS elements considered in the analysis.

$100\%$ is used to express the result in percentage.

Figure 3 displays the proximity of EHRMDS to each OS-EHRS in terms of overlapping and non-overlapping elements.

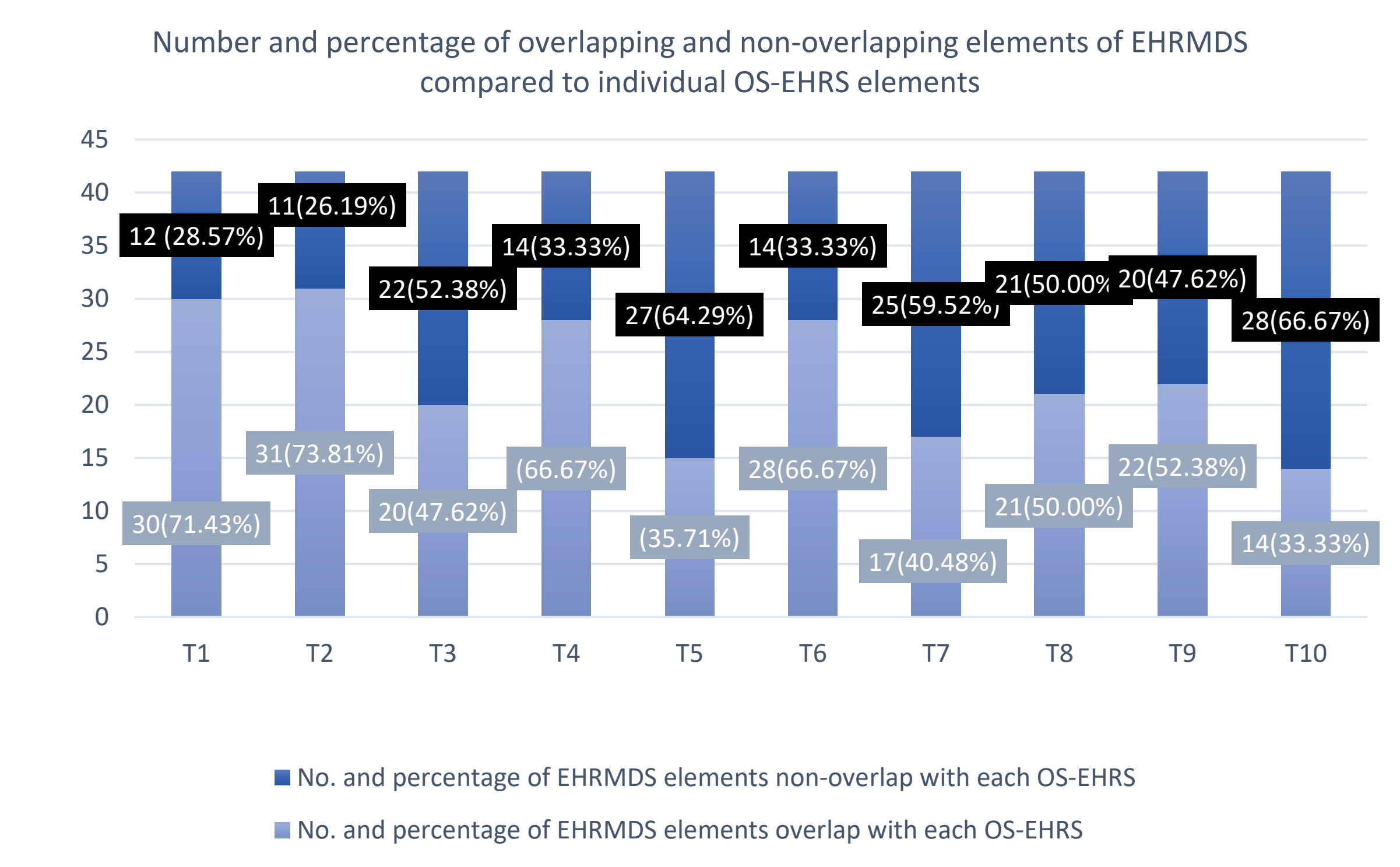

**Figure 3.** Displays the number of non-overlapping and overlapping elements of EHRMDS components in relation to each OS-EHRS.

The analysis reveals that EHRMDS was closest to T2, such as, OpenEMR. 31 out of 42 EHRMDS elements (73.81%) are present in T2, while the remaining 11 elements (26.19%) are absent. In contrast, Openclinic (T10) exhibits the lowest percentage of EHRMDS elements at 33.33%. The analysis indicates the existence of clinical elements that are included in EHR tools but absent in EHRMDS. This indicates that there is a need to develop an extended EHRMDS that can incorporate these missing elements. Therefore, in the following step, the authors develop an extended EHR schema. This step is crucial as it will enable EHRS to be more comprehensive in terms of clinical metadata coverage and enhance their overall utility, especially in the Indian healthcare system.

### 5.3. *Phase 3: Design an extended schema*

This phase involves the development of the extended EHR metadata, which expands upon the existing EHRMDS. The objective of this stage is to enhance the clinical metadata collection by incorporating components from OS-EHRS that were absent in EHRMDS. The authors analyse the absent components of each OS-EHRS to compile a distinct listing of elements. The list is merged with EHRMDS to create an extended schema.

Table 7 provides a summary of the count of elements in each tool and their presence or absence in EHRMDS. For instance, T1 (75Health) includes 48 elements, with 30 present in EHRMDS and 18 absent. The extended schema integrates these additional elements sourced from OS-EHRS. The formulas for calculating the present (P) and absent (A) elements are as follows:

*P = T - A (i.e., the count of elements present equals the total count of elements minus the count of absent elements).*

*A = T - P (i.e., the count of absent elements equals the total count of elements minus the count of present elements).*

Where:

*P: Represents the count of OS-EHRS elements present in EHRMDS.*

*A: Represents the count of OS-EHRS elements absent in EHRMDS.*

*T: Represents the total count of elements in OS-EHRS.*

**Table 7.** Count of Elements in Each OS-EHRS Present and Absent in EHRMDS

| | **T1 (48)** | **T2 (41)** | **T3 (28)** | **T4 (49)** | **T5 (26)** | **T6 (38)** | **T7 (28)** | **T8 (31)** | **T9 (38)** | **T10 (21)** |
|---|---|---|---|---|---|---|---|---|---|---|
| **# of elements present in EHRMDS** | 30 | 31 | 20 | 28 | 15 | 28 | 17 | 21 | 22 | 14 |
| **# of elements absent in EHRMDS** | 18 | 10 | 8 | 21 | 11 | 10 | 11 | 10 | 16 | 7 |

The final version of the extended metadata is shown in Table 10, which includes 89 elements, including 42 already-existing EHRMDS elements and forty seven new elements obtained from OS-EHRS systems. The final column displays the UMLS CUI IDs for the expanded list of elements, with 47 distinct elements highlighted in bold. By expanding the EHRMDS, a more comprehensive set of clinical metadata elements adapted from different EHR systems can be captured, making it more suitable for healthcare metadata coverage in India. The extended mapping metadata table including the terms from SNOMED CT[6] (Systematized Nomenclature of Medicine - Clinical Terms) [30] and UMLS CUI [31] and their Ids are provided in Table 10, Appendix.

## 6. Findings

This study systematically analyzes the closeness between the EHRMDS and OS-EHRS to evaluate their interoperability and metadata alignment. The results are derived from a quantitative closeness analysis that involved a crosswalk comparison of metadata elements, semantic mapping, and statistical evaluation. The key findings of this study are presented in the following subsections.

### 6.1. *Closeness Analysis of OS-EHRS to EHRMDS*

The study examined 10 widely used OS-EHRS and compared their metadata elements with the 42 core elements of EHRMDS. The closeness percentage was computed by identifying the proportion of overlapping metadata elements between each OS-EHRS and EHRMDS. The results indicate that OpenEMR exhibited the highest degree of compatibility with EHRMDS, covering 31 out of 42 elements (73.81%). In contrast, OpenClinic demonstrated the lowest alignment, covering only 14 out of 42 elements (33.33%).

[6] https://www.snomed.org/

The metadata alignment analysis for the 10 OS-EHRS is presented in Table 8, which ranks systems based on their Closeness Percentage (high to low) with EHRMDS. The table includes Gap Analysis (number of elements not matched: Total Metadata Elements - Elements Matched), Gap Percentage (percentage of elements not matched: 100% - Closeness Percentage), and Category (qualitative classification: High Alignment for Closeness Percentage ≥ 70%, Moderate Alignment for 50% ≤ Closeness Percentage < 70%, and Low Alignment for Closeness Percentage < 50%), providing a comprehensive view of alignment and areas for improvement.

**Table 8.** Closeness Percentage of OS-EHRS to EHRMDS

| Rank | OS-EHRS | Total Metadata Elements | Elements Matched with EHRMDS | Closeness Percentage (%) High ↑ to Low ↓ | Gap Analysis | Gap Percentage (%) | Category |
|---|---|---|---|---|---|---|---|
| **1** | **OpenEMR (T2)** | 41 | 31 | 73.81% | 10 | 26.19% | High Alignment |
| **2** | **NOSH EMR (T6)** | 38 | 28 | 66.67% | 10 | 33.33% | Moderate Alignment |
| **3** | **75Health (T1)** | 48 | 30 | 62.50% | 18 | 37.50% | Moderate Alignment |
| **4** | **Solismed (T4)** | 49 | 28 | 57.14% | 21 | 42.86% | Moderate Alignment |
| **5** | **OneTouchEMR (T9)** | 38 | 22 | 52.38% | 16 | 47.62% | Moderate Alignment |
| **6** | **GNU Health (T8)** | 31 | 21 | 50.00% | 10 | 50.00% | Moderate Alignment |
| **7** | **OpenMRS (T3)** | 28 | 20 | 47.62% | 8 | 52.38% | Moderate Alignment |
| **8** | **FreeHealth (T7)** | 28 | 17 | 40.48% | 11 | 59.52% | Low Alignment |
| **9** | **GNUMed (T5)** | 26 | 15 | 35.71% | 11 | 64.29% | Low Alignment |
| **10** | **OpenClinic (T10)** | 21 | 14 | 33.33% | 7 | 66.67% | Low Alignment |

$$Average\ Closeness = \frac{\sum(Closeness\ Percentage\ of\ each\ OS{-}EHRS)}{Total\ number\ of\ OS{-}EHRS} = 51.96\ \%$$

Standard deviation, $\sigma = \sqrt{\frac{1}{n}\sum(x_i - \bar{x})^2}$ = 14.2 % [Where, $x_i$ are the individual closeness percentages, $\bar{x}$= 52.96 is the compound average closeness, $n$ = 10 (total OS-EHRS)]

The computed average closeness percentage across the 10 OS-EHRS is 51.96%. The findings suggest that OpenEMR is the most compatible OS-EHRS for Indian healthcare needs, based on its high degree of metadata alignment with EHRMDS. However, there are still 26.19% missing elements in OpenEMR that are essential for full compliance with EHRMDS. RQ1 is addressed by quantifying the extent of metadata alignment, revealing an average closeness of 51.96% across OS-EHRS. RQ2 is answered by identifying OpenEMR as the most compatible OS-EHRS, achieving 73.81% alignment with EHRMDS.

### *6.2. Identification of Missing Metadata Elements in EHRMDS*

During the crosswalk analysis, 47 metadata elements were identified in OS-EHRS that were not present in EHRMDS. These missing elements primarily belong to the following categories:

i. *Advanced clinical parameters (ACP):* Certain OS-EHRS contained metadata for parameters such as genetic history, advanced diagnostic imaging data, and specific disease progression markers, which were missing from EHRMDS.
ii. *Administrative and operational data (AOD):* Several OS-EHRS included fields related to appointment scheduling, referral management, and billing, which were not considered in the EHRMDS framework.
iii. *Patient-reported data (PRD):* Features like lifestyle habits, dietary preferences, mental health assessments, and social determinants of health were more prevalent in OS-EHRS but absent in EHRMDS.

iv. *Interoperability and structured terminologies (IST):* Some OS-EHRS contained metadata elements mapped directly to international standards such as FHIR [32] and LOINC [33] , which were not part of EHRMDS.

Table 9 presents a summary of key metadata elements found in OS-EHRS but missing from EHRMDS.

**Table 9.** Missing Metadata Elements in EHRMDS Identified from OS-EHRS

| Category Abbrv. | Example Missing Elements |
|---|---|
| ACP | Genetic history, diagnostic imaging, disease progression markers |
| AOD | Appointment scheduling, referral tracking, billing codes |
| PRD | Lifestyle habits, dietary intake, social determinants of health |
| IST | FHIR mappings, LOINC codes, CDA-based structured data |

These findings indicate that EHRMDS requires an extension to better accommodate additional healthcare metadata elements that are already implemented in OS-EHRS. RQ3 is addressed by identifying the missing metadata elements in EHRMDS and their potential impact on interoperability. The findings suggest that EHRMDS requires expansion to include these elements for better alignment with global standards and improved data exchange.

### *6.3. Implications for Metadata Standardization and Interoperability*

The study's findings highlight critical gaps in EHR metadata standardization and interoperability. The lack of certain metadata elements in EHRMDS could lead to challenges in integrating OS-EHRS with national healthcare initiatives. The key takeaways from this analysis include:

i. *Need for Standardized Metadata Expansion:* Incorporating additional metadata elements into EHRMDS can enhance its applicability to real-world healthcare scenarios and improve data completeness.
ii. *Interoperability Challenges:* The variation in metadata coverage across different OS-EHRS suggests that healthcare organizations need the integrated frameworks to ensure seamless data exchange.
iii. *Policy Recommendations:* The results suggest that Indian healthcare policymakers should consider revising EHRMDS to incorporate missing metadata elements and align with global standards such as SNOMED-CT and UMLS.

This section addresses RQ4 and RQ5 by explaining how metadata gaps influence interoperability and standardization. RQ4 is answered by demonstrating how the quantitative closeness analysis (including gap analysis and standard deviation) can guide the selection of the most suitable OS-EHRS for the Indian healthcare system. For instance, OpenEMR emerges as the top choice due to its high alignment with EHRMDS. RQ5 is addressed by highlighting the implications of metadata gaps for standardization and interoperability. The findings emphasize the need for policy revisions and alignment with global standards to improve EHR systems' effectiveness.

### *6.4. Statistical Insights*

To provide a clearer understanding of metadata distribution and alignment across different OS-EHRS, Figure 3 illustrates the proportion of overlapping elements and non-overlapping elements of individual OS-EHRS in relation to EHRMDS.

The statistical analysis reveals:

- The average closeness percentage across all 10 OS-EHRS is 51.96%, indicating moderate compatibility with EHRMDS (Table 8).
- Three OS-EHRS (OpenEMR, 75Health, and NOSH EMR) demonstrated above 60% alignment, while two OS-EHRS (OpenClinic and GNUMed) had less than 40% alignment.
- The standard deviation in closeness percentage is 14.2%, reflecting the significant variability in metadata structure among OS-EHRS. This high standard deviation suggests that OS-EHRS do not have uniform metadata structures, leading to significant variability in how well they align with EHRMDS. Some are well-structured for interoperability, while others lack key metadata elements. These statistical insights reinforce the findings for RQ1 (extent of alignment) and RQ4 (selection of suitable OS-EHRS), providing a data-driven basis for decision-making.

## 7. Conclusion and future directions

The study systematically analyzed the similarity of EHR metadata and developed an extended EHR schema. Adopting a quantitative approach, it compared metadata elements from various OS-EHRS with the EHRMDS and extended the framework accordingly. The extended metadata is further aligned with healthcare information exchange standards such as UMLS and SNOMED-CT, offering insights into the compatibility of existing OS-EHRS with EHRMDS.

The findings highlight OpenEMR as the most aligned OS-EHRS with EHRMDS while revealing persistent metadata gaps across all systems. The identification of 47 missing elements in EHRMDS underscores the necessity of extending the metadata framework to meet modern clinical, administrative, and interoperability requirements.

Despite its contributions, this study has certain limitations. The analysis was limited to a small set of OS-EHRS, excluding commercial EHR systems that are widely used in healthcare settings. Future research should expand this evaluation to encompass proprietary EHR platforms to provide a more comprehensive assessment of metadata compatibility. Additionally, while this study focused on metadata closeness analysis between OS-EHRS and EHRMDS, it did not assess interoperability with other global healthcare information standards such as HL7 FHIR [34]. Evaluating the compatibility of OS-EHRS with FHIR and other frameworks would offer a more holistic understanding of metadata standardization and healthcare data exchange [32, 35].

The proposed extended EHR metadata require further validation in real-world clinical settings to assess their practical utility in enhancing data interoperability. Future research should also include a comprehensive mapping of the extended metadata list with FHIR terminologies to enable seamless data exchange.

This study contributes to the advancement of open-source EHR metadata and supports the widespread adoption of structured, interoperable health records. The systematic approach to closeness analysis can be particularly beneficial for countries like India, where the Ministry of Health and Family Welfare has recommended EHR standards and EHRMDS. The proposed extended schema lays the foundation for further studies in medical ontology development, knowledge base frameworks, and interoperability of healthcare data.

### Abbreviations

ACP, Advanced Clinical Parameters; AM, Archetype Model; AOD, Administrative and Operational Data; ASTM, American Society for Testing and Materials; BP, Blood Pressure; CCR, Continuity of Care Record; CDA, Clinical Document Architecture; CDSS, Clinical Decision Support System; CHITS, Community Health Information Tracking System; CUI, Concept Unique Identifier; DMP, Dossier Medical Personnel; EHR, Electronic Health Record; EHRMDS, Electronic Health Record Minimum Data Set; EHRS, Electronic Health Record System; FHIR, Fast Healthcare Interoperability Resources; LBF, Literature-Based Frequency; GNU, GNU's Not Unix; HIS, Hospital Information System; HL7, Health Level 7; HITECH, Health Information Technology for Economic and Clinical Health; ICD, International Classification of Diseases; ICPM, International Classification of Procedures in Medicine; LOINC, Logical Observation Identifiers Names and Codes; MMS, Massachusetts Medical Society; MoHFW, Ministry of Health and Family Welfare; MU, Meaningful Use; NLP, Natural Language Processing; OAE, Online Accessibility for Evaluation; OS-EHRS, Open-Source Electronic Health Record Systems; PCMH, Patient-Centered Medical Home; PRD, Patient-Reported Data; PRISMA, Preferred Reporting Items for Systematic Reviews and Meta-Analyses; RM, Reference Model; SNOMED-CT, Systematized Nomenclature of Medicine – Clinical Terms; UMLS, Unified Medical Language System; UHID, Unique Health Identifier.

# Appendix

**Table 10.**Shows crosswalk between EHRMDS and OS-EHRS. It also provides the extended EHRMDS i.e., the extended EHR schema

| EHRMDS | OS-EHRS | | | | | | | | | | EHRMDS-ext. | UMLS CUI |
|---|---|---|---|---|---|---|---|---|---|---|---|---|
| | 75Health | OpenEMR | OpenMRS | Solismed | GNUMed | NoshEMR | Freehealth | GNUHealth | Onetouchemr | Openclinic | | |
| Encounter Type | Encounter type | Encounter type | Encounter type | Encounter | Encounter type | Encounter | Encounter type | Encounter type | Encounter type | Encounter | Encounter Type | C0586016 |
| Encounter Number | Encounter no | Patient Encounter no | Patient Encounter number | Encounter number | Encounter no | Encounter Number | | Encounter Number | | | Encounter Number | C4086434 |
| Encounter Date & Time | Encounter Date | Encounter Date | Encounter Date | | Encounter Date | | | Encounter Date | | | Encounter Date | C4087739 |
| Reason for Visit | Visit Reasons | Visit Reason | | Reason | | Reason | | Indication | | Reason | Reason for Visit | C1704447 |
| | Physical Examination | | | Physical Exam | | | | | | | **Physical Examination** | C0031809 |
| | | | | Review of Systems | | | | | | | **Review of Systems** | C0489633 |
| Present History | | Present History | History and Examination | HPI | | Present Medical History | | | Present Medical History | | Present History | C1827598 |
| | | | | | | | | | Conservative Therapy | | **Conservative Therapy** | C0459914 |
| | | | | Surgical History | | Surgical History | Patient Overview: Surgical | | Surgical History | | **Surgical History** | C0455610 |
| Past History | | General History | | Past Medical History | | | Patient Overview: Past | | | | Past History | C0455458 |
| Personal History | | Life Style | | | | | Patient Overview: Personal | Individuals, lifestyle | | Personal Antecedents | Personal History | C0585172 |
| Family History | Family Health | Family History | | Family History | | Family History | Patient Overview: Family | Families | Family History | Family Antecedents | Family History | C0241889 |
| Menstrual & Obstetric History | | | Obstetrics Gynecology | | | | patient overview: obstetric | OB/GYN | Last Menstrual | | Menstrual & Obstetric History | C0425963 |
| Socio-economic Status | Social History | | | Social History | | social history | | Socioeconomics | Social History | Social Data | Social History | C0424945 |
| Immunization History | Vaccine & Immunization | | | Immunizations | | Immunizations | | vaccine | Imm/ Injections | | Immunization History | C0552506 |
| Clinical Exam Vitals Systolic BP | Patient Health Record vital: Blood Pressure | BP Systolic | Blood Pressure | Blood Pressure | BP Systolic | Systolic | BP | | Blood Pressure | | Systolic Blood Pressure | C0871470 |
| Clinical Exam Vitals Diastolic BP | Patient Health Record vital: Blood Pressure | BP Diastolic | | | BP Diastolic | Diastolic | | | | | Diastolic Blood Pressure | C0428883 |
| Clinical Exam Pulse Rate | Patient Health Record vital: Heart rate/ pulse | Pulse | Pulse | Pulse | Heart Rate | Pulse | | | Pulse | | Pulse Rate | C0577836 |
| Clinical Exam Temperature (°C) | Patient Health Record vital: Temperature | Temperature | Temperature (C) | Temperature | Body Temp (degree c) | Temperature | T | Temperature | Temperature | | Temperature | C0039476 |
| Clinical Exam Temperature Source | | Temp Location | | | Rectal temp | | | | | | Temperature Source | C0204688 |
| Clinical Exam Respiration Rate | Patient Health Record vital: Respiratory rate | Respiration | Respiratory rate | Respiratory | RR | RR | | | Respiratory Rate | | **Respiratory Rate** | C0231832 |
| | | | | Breathing Pattern | | | | | | | Breathing Pattern | C0517967 |
| Clinical Exam Height (cms) | Patient Health Record vital: height & Patient Health Record vital: Height at > 25 | Height/Length | Height (cm) | Height | Height | Height | Height | Height | Height | Height | Height | C0005890 |
| Clinical Exam Weight (kgs) | Patient Health Record vital: weight | Weight | Weight (kg) | Weight | Weight | Weight | Weight | Weight | Weight | Weight | Weight | C0005910 |
| Blood Group | | | | | | | | | | | Blood Group | C0600103 |
| Clinical Exam Observation | | | Latest observation | Observation | | | | | | Observation | Clinical Exam Observation | C1274016 |
| Investigation Results | | | | Lab Result | Lab Results | Results | Consultation reports Lab test report | | Labs | | Investigation Results | C0587081 |
| Clinical Summary | Problems | Medical Problems | | Problem list | | Problem list | | Problem | Problem list | Medical Problem | Clinical Summary | C0033213 |
| | | | Health Maintenance therapy | Health Maintenance summary | | | | | | | **Health Trend Summary** | C0677908 |
| Diagnosis Type | Diagnosis Type | Diagnosis Type | | Diagnosis Type | | | Diagnosis Type | | | | Diagnosis Type | C0332131 |
| Diagnosis Code Name | Diagnosis Code Name | Diagnosis Code Name | | Diagnosis Code Name | | Diagnosis Code Name | | | | Diagnosis Code Name | Diagnosis Code Name | C2985803 |
| Diagnosis Code | Diagnosis Code | Diagnosis Code | | Diagnosis Code | | Diagnosis Code | | | | Diagnosis Code | Diagnosis Code | C1550350 |
| Diagnosis (Description) | Diagnosis (Description) | Diagnosis (Description) | Diagnosis | Diagnosis (Description) | Diagnosis | Diagnosis (Description) | | | Diagnosis | Diagnosis Description | Diagnosis (Description) | C0011900 |
| Treatment Plan Investigations | | | | | | | | | | | Treatment Plan Investigations | |
| Treatment Plan Medication | Plan of Treatment | Treatment Plan | | Treatment | Treatment s | Medical Action Plan | | Treatments | Health Maintenance Plans | | Treatment Plan Medication | C4545837 |
| Treatment Plan Procedure | Procedure | Procedure | | Procedures | | | Programs | Procedures | Procedure | | Treatment Plan Procedure | C0237403 |
| Treatment Plan Referral | Referral | Referral | | Referrals | | | | | | | Treatment Plan Referral | C0814457 |
| Other Treatment Plan Type | | | | | | | | | | | Other Treatment Plan Type | |
| Other Treatment Plan Details | | | | | | | | | | | Other Treatment Plan Details | |
| Current Clinical Status | | | Conditions: Active Conditions: History Of Conditions: Inactive | | Conditions | Condition | | | | | Current Clinical Status | C3899485 |
| Medication Name | Medication: Medicine name | Medication | Medication | Medication List: Drug | | Medication RXNorm | Long-term Medication | Medicaments | Medication RxNorm | Drugs | Medication Name | C2360065 |
| Drug Code | Drug code | Drug Code | Drug Code | | | Drug code | | | Drug code | | Drug Code | C54185 |
| Strength | Medication: Medicine Strength | Strength | Strength | | | Strength | Strength | Strength | Strength | | Strength | C1705922 |
| Dose | Dose | Dose | Dose | Medication List: Direction Dose | | Dosage | Dose | Dose | | | Dose | C0178602 |
| Route | Route | Route | Route | | | Route | routes | Administration Route | | | Route | C0013153 |
| | Medication: Instructions | | | Narration & Instruction | | Special Instructions | | | | | Special Instructions | C1442085 |
| Frequency | Frequency | Frequency | | | | Frequency | Duration | Frequency | | | Frequency | C89081 |
| | Medication: Refills | Refills | | Refills | | | refill | Refills | Refill Allowed | | Refills | C4289665 |
| | Medication: Start Date | | | | | | Date and time | Start: Date and time | Start Date | | **Medication: Start Date** | C5141805 |
| | Medication: End Date | | | | | | limits | End: Date and time | End Date | | **Medication: End Date** | C5141806 |
| | Medication: Company Name | | | Medication List: Package Description | | | | | | | **Medication: Company Name** | C0815286 |
| Allergy Status | Allergy: Status | Allergies | | Allergy status | | Allergies | | Patients: Patient allergies and Critical Information | Allergies: Status | Allergy | Allergy: Status | C4521222 |
| Allergy History | Allergy: Allergen Name | | Allergies | allergy: agent | | | | | Allergies: Agent | | Allergy: Allergen Name | C0002092 |
| | Allergy: Type | | Allergy type | Allergy: type | | | Drug Allergies | | Allergies: Type | Allergy type | **Allergy: Type** | C1550403 |
| | Allergy: Reactions | | Reactions | Allergy: reactions | | Reaction | | | Allergies: Reaction | | **Allergy: Reactions** | C1527304 |
| | Allergy: Severity | | severity | Allergy: severity | | Severity | | | Severity | | **Allergy: Severity** | C1550404 |
| | Allergy: From Date | | Allergy Date | Allergy:date recorder | | | | | Start Date, End Date | | **Allergy: From Date** | C2209280 |
| | | | | | | | | | Allergies: Source | | **Allergies: Source** | C1690571 |
| | | | | | | Substance or Medication | | | | | **Substance or Medication** | C3266231 |
| | | | | Allergy: conditions | | | | | | | **Allergy: conditions** | C0851444 |
| | Patient Health Record vital: Oxygen Saturation | Oxygen Saturation | Blood oxygen saturation | SpO2 | SPO2 | | | | SpO2 | Spo2 | **SpO2** | C0513686 |
| | Blood Glucose | | | | | | | | | | **Patient Health Record vital: Glucose by Glucometer** | C3669205 |
| | Patient Health Record vital: BMI | BMI | (Calculated) BMI | BMI | BMI | BMI Status | BMI | | BMI | BMI | **BMI** | C0578022 |
| | | Head Circumferen ce | | Head Circumference | | Head Circumferen ce | Head Circumference | | Head Circ | | **Head Circumference** | C0262499 |
| | | Waist Circumferen ce | | Waist Size | Waist Circum | | | | Waist | | **Waist Circumference** | C0455829 |
| | Test order | Past Test Order | | Lab Order Status | | Orders | Lab Order Fulfillments | | | | **Lab Order** | C4302923 |
| | Implantation process | | | | | | | | | | **Implantation procedure** | C0021107 |
| | Goals | | | | | | | | | | **Goals** | C0557971 |
| | | Surgeries | | | | | | Surgeries | | | **Surgeries** | C0543467 |
| | | Dental treatment | | | | | | | | | **Dental procedure** | C0011331 |
| | | | | | Posture | | | | | | **Posture** | C1262869 |
| | | Diagnostic Imaging | | | Pacs orders | | | Imaging | | Imaging | **Imaging** | C0011923 |
| | | | | Pathology Order | Pathology Orders | | | | | | **Pathology order** | C4302922 |
| | | | | Radiology Order | Radiology Orders | | | | | | **Radiology Order** | C4302924 |
| | | | | Radiology Result | Radiology | | | | Radiology | Radiology | **Radiology** | C1290916 |
| | | | | | Disposition | | | | | | **Disposition** | C0743223 |
| | | | | | followup | | | | | | **Follow-up** | C0589120 |
| | | | | | | sensitive level | | | | | **sensitive level** | C1455667 |
| | | | | | | Supplement | | | | Supplements | **Supplement** | C0242295 |
| | | | | | Urine Sugar | | | | | | **Urine Sugar** | C1456823 |
| | | | | | | | Drug Intolerance | | | | **Drug Intolerance** | C0277585 |
| | | | | | | | | Pediatrics History | | | **Pediatrics History** | C0587599 |
| | | | | | | | | Pediatrics Growth Charts | | | **Pediatrics Growth Charts** | C2718056 |
| | | | | | | | | Recreational Drugs | | | **Recreational Drugs** | C1318816 |
| | | | | | | | | Medical Specialities | | | **Medical Specialities** | C0037778 |
| | | | | | | | | Pages of Life: Genetics | | | **Genetics** | C3887703 |
| | | | | | | | Patient overview: Risk factors | | | | **Risk factors** | C0035648 |
| | Advance Directive: Assessment | | | | | | | | Advance Directive | | **Advance Directive** | C0587820 |
| | Medication: Directions for use or SIG CODE | SIG code | SIG code | | | | SIG code | | | SIG | **Medication: Directions for use or SIG CODE** | C3476380 |